\documentclass[prl,aps,twocolumn,showpacs]{revtex4}
\usepackage{amssymb}
\usepackage{graphicx}
\usepackage{dcolumn}
\usepackage{bm}
\usepackage{amsmath}
\usepackage{float}
\begin{document}

\title{Excitation of multiple wakefields by short laser pulses in dense plasmas}

\author{P. K. Shukla}
\affiliation{Institut f\"ur Theoretische Physik IV, Ruhr--Universit\"at Bochum, 
D-44780 Bochum, Germany; Department of Physics, Ume{\aa} University, SE--901 87 Ume{\aa}, 
Sweden; Scottish Universities Physics Alliance, Department of Physics, 
University of Strathclyde, Glasgow G4 0NG, Scotland, United Kingdom}

\author{Gert Brodin}
\affiliation{Department of Physics, Ume{\aa} University, SE--901 87
Ume{\aa}, Sweden}

\author{Mattias Marklund}
\affiliation{Department of Physics, Ume{\aa} University, SE--901 87
Ume{\aa}, Sweden}

\author{Lennart Stenflo}
\affiliation{Department of Physics, Ume{\aa} University, SE--901 87
Ume{\aa}, Sweden}

\received{20 March 2009}

\begin{abstract}
We present a theoretical investigation of the excitation of multiple electrostatic wakefields 
by the ponderomotive force of a short electromagnetic pulse propagating through a dense 
plasma. It is found that the inclusion of the quantum statistical pressure and 
quantum electron tunneling effects can qualitatively change the classical 
behavior of the wakefield. In addition to the well known plasma oscillation 
wakefield, with a wavelength of the order of the electron skin depth ($\lambda_e= c/\omega_{pe}$,
which in a dense plasma is of the order of several nanometers, where $c$ is the speed 
of light in vacuum and $\omega_{pe}$ is the electron plasma frequency), wakefields
in dense plasmas with a shorter wavelength (in comparison with $\lambda_e$) are also 
excited. The wakefields can trap electrons and accelerate them to extremely high energies
over nanoscales. 
\end{abstract}

\pacs{52.35.Fp, 52.35.Mw, 52.38.Kd}

\maketitle

Three decades ago, Tajima and Dawson \cite{Tajima-1979} demonstrated that intense laser 
pulses can efficiently generate electron plasma waves (EPWs) in their wake as they 
travel through a low density plasma. Physically, the ponderomotive force \cite{Shukla-1986} of an intense 
laser pulse pushes electrons locally in plasmas, which in turn oscillate at the electron plasma frequency 
with respect to the neutralizing background of immobile positive ions. The displacement of the electrons
within the plasma gives rise to electric fields, which can be much larger than any fields 
possible in a non-ionized material. An electron beam can surf on the electric field of a plasma
wave picking up energy from the EPWs just as a surfer picks up energy from a water wave in the 
ocean \cite{Mendonca-2001}. The idea of Tajima and Dawson has now been experimentally verified  
\cite{Mangels-2004, Leemans-2006,Joshi-2006,Blumenfeld-2007,Bingham-2007,Leemans-2009}.

Recently, there has been a great deal of interest in investigating the properties of 
high-energy density plasmas that are created by high intensity laser pulses. To probe dense
matter, such as those in the interior of white dwarf stars and Jovian planets,
powerful laser-produced x ray sources have been developed. They produce monoenergetic line
radiation capable of penetrating through dense and compressed materials at solid densities
and above \cite{Glenzer-2007}. The x-ray measurement techniques are indicative of 
a dense Fermi-degenerate plasma state in laboratories.  In a dense Fermi plasma, the 
electron degeneracy leads to a consideration of the Fermi-Dirac electron distribution 
and electron quantum tunneling through the quantum Bohm 
potential \cite{Gardner-1995,Manfredi-2005,Shukla-2006}. Furthermore, there are also 
the spin force and spin magnetized electron current due to electron-$1/2$ 
spin in dense magnetoplasmas \cite{Marklund-2007,Shukla-2009}. The quantum statistical
pressure, the quantum Bohm force and the quantum spin force drastically affect
the electron dynamics, and therefore one encounters numerous novel collective 
interactions in dense quantum plasmas. Specifically, it should be stressed that 
the quantum Bohm force effect, arising from the finite width of the electron wave 
function, gives rise to dispersion of EPWs at nanoscales, which has important 
consequences to localized EPW structures \cite{Shukla-2006,Shukla-2007} and 
plasmonic turbulence \cite{Shaikh-2007}.

In this Letter, we present a theoretical investigation of the multiple EPW (wakefield) 
excitation by the ponderomotive force of a short laser pulse, accounting for the 
quantum statistical pressure and quantum Bohm force effects in the EPW dynamics.
For our purposes, we shall use the quantum Madelung fluid equations \cite{Manfredi-2005},
which are composed of the electron continuity and electron momentum equations,
together with the Poisson equation, and derive the EPW (wakefield) equation in 
the presence of the radiation pressure. Chosing a specific form for the laser envelope, 
we solve the EPW equation analytically and numerically. We find that, due to the quantum Bohm 
force, multiple wakefields at nanoscales appear in dense quantum plasmas. 
In addition to the well known wakefield, with a wavelength of the order of the electron 
skin depth, a short wavelength wakefield is also excited, with a scale length approaching 
the Compton wavelength. It turns out that for the case of a laser pulse in the optical regime, 
the short scale wakefield is suppressed.  However, for short laser pulse lengths and/or high 
density plasmas, the energy density of the short scale wakefield may be comparable to that 
of the long wavelength wakefield. The consequences of our results are discussed.

We consider the propagation of a high-frequency laser pulse, with the vector
potential $\mathbf{A}=\widetilde{\mathbf{A}}\exp (ikx- i \omega t)$ $+\mathrm{%
c.c.}$, in an unmagnetized dense plasma. Here $\mathrm{c.c.}$ stands
for the complex conjugate. The ponderomotive force of the high-frequency laser 
pulse drives longitudinal EPWs (wakefields) with a frequency much smaller
than $\omega $, but fast enough for the dynamics to take place on the
electron timescale. The ions form a neutralizing background in our dense plasma. 
The governing equations for the wakefields are then the electron continuity equation

\begin{equation}
\frac{\partial n_1}{\partial t}+\nabla \cdot \left( n_0\mathbf{v}\right) =0,
\label{Eq. cont}
\end{equation}
the electron momentum equation

\begin{equation}
\frac{\partial \mathbf{v}}{\partial t}= \frac{e}{m}\nabla \Phi -\frac{e^{2}}{%
2m^{2}c^{2}}| \widetilde{\mathbf{A}}| ^{2}-\frac{V_{F}^{2}}{n_{0}}%
\nabla n_{1}+\frac{\hbar ^{2}}{4m^{2}}\nabla \nabla ^{2}n_{1},
\label{Eq-moment}
\end{equation}
and
the Poisson equation

\begin{equation}
n_{1}=\frac{\nabla ^{2}\Phi }{4\pi e},  
\label{Eq: Poisson}
\end{equation}
where $n_1 $ is the electron density perturbation in the equilibrium
value $n_0$, $\mathbf{v}$ is the electron fluid velocity perturbation,
$\Phi$ is the wakefield potential, $e$ is the magnitude of the electron charge,
$m$ is the electron rest mass, $V_{F}=(2\pi \hbar /\sqrt{3} m)(3n_{0}/8\pi )^{1/3}$ 
is the Fermi speed, and $\hbar $ is the Planck constant divided by $2\pi$. We have 
thus assumed that the Fermi electron pressure dominates over the electron thermal pressure, 
appropriate for a high density plasma of moderate or low electron temperature. 
Several comments are in order. The second term in the right-hand side of (2), which 
represents the light pressure or the laser ponderomotive force, comes from averaging 
(over the laser period) the advection and nonlinear Lorentz force in the electron 
equation of motion \cite{Shukla-1986}. The fourth term in the right-hand side of (2) is 
the quantum Bohm force involving quantum electron tunneling in a dense quantum plasma.

Combining Eqs. (1)-(3) we obtain the plasma wakefield equation in the presence 
of the light pressure in our dense plasma

\begin{equation}
\left[ \frac{\partial ^{2}}{\partial t^{2}}+\omega _{p}^{2}-v_{F}^{2}\nabla
^{2}+\frac{\hbar ^{2}}{4m^{2}}\nabla ^{4}\right] \Phi =\frac{\omega _{p}^{2}%
}{2c^{2}m}| \widetilde{\mathbf{A}}| ^{2},
\label{Eq: LF-general}
\end{equation}
where $\omega_{pe} =(4\pi n_0e^ 2/m_e)^ {1/2}$ is the electron plasma frequency.
In dense laboratory plasmas and in compact astrophysical objects, the latter is
in the x-ray regime.

Let us now consider the excitation of a one-dimensional wakefield by the high-frequency 
laser pulse that is propagating with the group velocity. Thus, we look for stationary 
solutions of (4) in a comoving frame. Letting $\xi =x-v_{g}t$, where $v_{g}$ is the 
group velocity, we obtain from Eq. (\ref{Eq: LF-general}) 

\begin{equation}
\left[ \frac{\hbar ^{2}}{4m^{2}}\frac{\partial ^{4}}{\partial \xi ^{4}}%
+\left( v_{g}^{2}-v_{F}^{2}\right) \frac{\partial ^{2}}{\partial \xi ^{2}}%
+\omega _{p}^{2}\right] \Phi =\frac{\omega _{p}^{2}}{2c^{2}m}| 
\widetilde{\mathbf{A}}| ^{2}.
\label{Eq: One dimension}
\end{equation}

A simple special case is found when $v_{g}^{2}=v_{F}^{2}$, which results in
exponentially damped wakefields. However, below we will consider the case
where we can have multiple oscillatory wakefields, which occur for plasmas
of moderate density (see the condition below). We start by investigating the
solutions to the left-hand side of Eq. (5) in the absence of the driving
laser field. Making the ansatz $\Phi \varpropto \exp (ik\xi )$,  we obtain
$^{4}-k_{a}^{2}k^{2}+k_{b}^{4}=0$,
where $k_{a}^{2}=4m^{2}\left( v_{g}^{2}-v_{F}^{2}\right) /\hbar ^{2}$ and $%
k_{b}^{4}=4m^{2}\omega _{p}^{2}/\hbar ^{2}$. This has the solution

\begin{equation*}
k^{2} = \frac{k_{a}^{2}}{2}\pm \sqrt{\frac{k_{a}^{4}}{4}-k_{b}^{4}}.
\end{equation*}
Unless the plasma density is very high (or $v_{g}^{2}\simeq v_{F}^{2}$), $\
k_{a}^{4}/4\gg k_{b}^{4}$, and the two solutions separate into oscillations
with very different scale lengths. Denoting the solutions with $k_{+}$ and $%
k_{-}$, we have for $k_{a}^{4}/4\gg k_{b}^{4}$%

\begin{eqnarray*}
k_{+}^{2} &\approx &k_{a}^{2}=4m^{2}\left( v_{g}^{2}-v_{F}^{2}\right) /\hbar
^{2} , \\
k_{-}^{2} &\approx &k_{b}^{4}/k_{a}^{2}=\omega _{p}^{2}/\left(
v_{g}^{2}-v_{F}^{2}\right) 
\end{eqnarray*}

The analysis below will be valid for arbitrary values of $k_{+}$ and $k_{-}$, 
however, as long as both modes are oscillatory, which holds whenever 
%
$v_{g} >v_{F} $
and
$m\left| v_{g}^{2}-v_{F}^{2}\right| > \hbar \omega _{p} $
apply. When the second inequality is a strong one, the oscillations at $k_{+}$
is close to the Compton scale (at least when $v_{g}^{2}-v_{F}^{2}\simeq c^{2}
$), and the $k_{-}$ oscillation is the standard plasma oscillation
wakefield, although somewhat modified by the inclusion of the Fermi
pressure. As a starting point, however, no separation in magnitude of the
two scales will be assumed. Firstly, we rewrite Eq. (\ref{Eq: One dimension})
as 

\begin{equation}
\left[ \frac{\partial ^{4}}{\partial \xi ^{4}}+k_{a}^{2}\frac{\partial ^{2}}{%
\partial \xi ^{2}}+k_{b}^{4}\right] \Phi =\eta | \widetilde{\mathbf{A}}%
| ^{2},
\label{Eq: Normalized}
\end{equation}
where $\eta =2m\omega _{p}^{2}/\hbar ^{2}c^{2}$. Eq. (\ref{Eq: Normalized})
can be integrated to yield 

\begin{equation}
\Phi =\frac{\eta }{(k_{+}^{2}-k_{-}^{2})}\int_{-\infty }^{\xi }\left[ \frac{1%
}{k_{+}}\sin [k_{+}(\xi -\xi ^{\prime })]-\frac{1}{k_{-}}\sin [k_{-}(\xi
-\xi ^{\prime })]\right] \eta | \widetilde{\mathbf{A}}| ^{2}(\xi
^{\prime })d\xi ^{\prime },  
\label{Eq: Green fnk result}
\end{equation}
where we have assumed as boundary condition that $\Phi $ is zero before the
arrival of the high-frequency laser pulse.

Next, we assume that the laser pulse profile is a Gaussian, viz. $| A|
^{2}=A_{0}^{2}\exp (-\xi ^{\prime 2}/L^{2})$. The energy density of the
high-frequency laser pulse is assumed to be high enough such that the 
high-frequency laser pulse is changing its shape on a longer timescale, 
as compared to the wakefield generation process. The simplest case is that 
of a very short pulse, i.e. when $L\ll k_{-}^{-1},k_{+}^{-1}$, when the wakefield 
after the laser pulse passage can be written as

$\Phi =\Phi _{+}\sin (k_{+}\xi +\varphi _{+})+\Phi _{-}\sin (k_{-}\xi
+\varphi _{-})$, 
where $\varphi _{+}$ and $\varphi _{-}$ are constant phase angles and the
amplitudes of the wakefields are proportional to the high-frequency laser 
pulse energy, and given by 

\begin{eqnarray*}
\Phi _{+} &=&\frac{\eta }{(k_{+}^{2}-k_{-}^{2})k_{+}}\int_{-\infty }^{\infty
}| \widetilde{\mathbf{A}}| ^{2}(\xi ^{\prime })d\xi ^{\prime } \\
\Phi _{-} &=&\frac{\eta }{(k_{+}^{2}-k_{-}^{2})k_{-}}\int_{-\infty }^{\infty
}| \widetilde{\mathbf{A}}| ^{2}(\xi ^{\prime })d\xi ^{\prime }.
\end{eqnarray*}

In general, the wakefield amplitudes after the laser pulse passage can be written as

\begin{equation*}
\Phi _{\pm }=\frac{\eta }{(k_{+}^{2}-k_{-}^{2})k_{_{\pm }}}\int_{-\infty
}^{\infty }\cos (k_{\pm }\xi )| \widetilde{\mathbf{A}}| ^{2}d\xi
^{\prime }
\end{equation*}
such that the electric field amplitudes $E_{\pm }$ of the different wakefields 
in the case of a Gaussian profile is given by 

\begin{equation}
E_{\pm }=\frac{\eta }{(k_{+}^{2}-k_{-}^{2})}W\exp (-k_{\pm }^{2}L^{2}/4)
\label{Short-wake field}
\end{equation}
where we have introduced $W=\int_{-\infty }^{\infty }| \widetilde{%
\mathbf{A}}| ^{2}(\xi ^{\prime })d\xi ^{\prime }$ proportional to the
high-frequency laser pulse energy. (Moreover, we have used $\int_{-\infty }^{\infty }$ $\cos
(y)\exp (-y^{2}/L^{2})dy=L\sqrt{\pi }e^{-L^{2}/4}$) Thus, we see that
efficient wakefield generation requires that the laser pulse length is not much
longer than the respective wakefield wavelength. If the Fermi speed is
much smaller than the speed of light in vacuum, and $\omega \gg \omega _{p}$,
significant excitation of the long wavelength wakefield requires that the laser pulse length
is not much longer than the electron skin depth, whereas significant excitation of the
short scale wakefield requires a laser pulse length not much longer than the
Compton wavelength. This condition of an extremely short pulse can be
relaxed to some extent in a high-density plasma where the Fermi speed
approaches the speed of light in vacuum, such as in a white dwarf star. For such
densities we may have significant excitation also of the short scale wakefield 
for laser pulse lengths $L\sim \hbar /m(v_{g}^{2}-v_{F}^{2})^{1/2}$ or
shorter. The full profile of the density perturbation $n_{1}$ induced by the
high-frequency laser pulse is calculated numerically from Eqs (\ref{Eq: Poisson})
and (\ref{Eq: Green fnk result}) for the  plasma number density $n_{0}\simeq 10^{28}%
\mathrm{cm}^{-3}$ (which is achievable in inertial fusion experiments) and a pulse
length $L$ of a few Compton wavelengths (i.e.\ $k_{+}L\simeq 7$). The result
is shown in Fig. 1. Due to the large separation in length-scales, the
quantum scale wakefield is hidden to some extent in the large scale plot.
However, the quantum scale oscillations are clearly shown in the zoom made in
the sub-panel of Fig 1. For a higher plasma number density $n_{0}\simeq 10^{30}%
\mathrm{cm}^{-3}$ (as relevant for e.g. white dwarf stars) and a slightly
shorter pulse length, the quantum scale oscillations can be directly seen,
as pictured in Fig 2 (pulse length $k_{+}L\simeq 4$) and Fig 3. (pulse
length $k_{+}L\simeq 3)$.

To summarize, we have presented a theoretical study of the EPW excitation 
by short laser pulses in a dense plasma. For this purpose, we have 
used the electron continuity and electron momentum equation including the quantum statistical
pressure, the quantum Bohm force, and the light pressure, together with the Poisson
equation to derive the driven EPW equation. The latter shows that the 
quantum force produces an additional EPW dispersion, besides that coming from
the quantum statistical pressure.  Our results reveal that the
quantum Bohm force provides the possibility of a short wavelength (in 
comparison with the electron skin depth) wakefield, in addition to the standard 
plasma oscillation wakefield, with a wavelength of the order of the electron skin depth. 
For current experiments with the laser pulses in the optical regime, the short scale wakefield 
is suppressed. However, for very short laser pulses, of the order of a few Compton wavelengths, 
or for high density plasmas where the Fermi speed approaches the speed of light in vacuum, 
the energy density of the short scale wakefield is comparable to that of the long
wavelength wakefield. The energy loss of the laser pulse is then
correspondingly higher. The laser excited wakefields can trap electrons and accelerate them
to high energies at nanoscales in dense plasmas, such as those in the next generation 
intense laser-solid density plasma experiments and in compact astrophysical objects 
\cite{Harding-2006} (e.g.\ the interior of white dwarf stars and the planet Jupiter).  
Finally, the present investigation can be readily generalized by including an external 
magnetic field, in which case the propagation of both laser pulses and wakefields will be 
greatly affected.

\acknowledgments
This work was supported by the European Research Council under Contract No.\
204059-QPQV, and by the Swedish Research Council under Contract No.\ 2007-4422.

\begin{figure}
\includegraphics[width=.9\columnwidth]{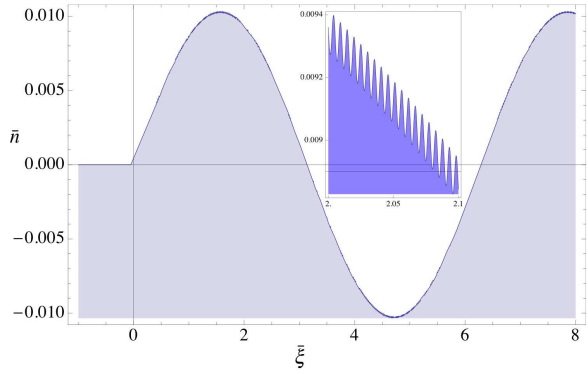}
\caption{The total normalized plasma density wakefield for $n_{0}\simeq
10^{28}\mathrm{cm}^{-3}$ and a Gaussian laser pulse profile $\mathbf{A}%
_{0}^{2}\exp (-\xi ^{2}/L^{2})$with $k_{+}L\simeq 7$. A zoom is made in the
sub-panel to make the quantum scale oscillations clear. The normalizations
used are $\overline{\xi }=\omega _{p}\xi /c$ and $\overline{n}=\varepsilon
_{0}\eta \mathbf{A}_{0}^{2}n/q_{e}$.}
\end{figure}

\begin{figure}
\includegraphics[width=.9\columnwidth]{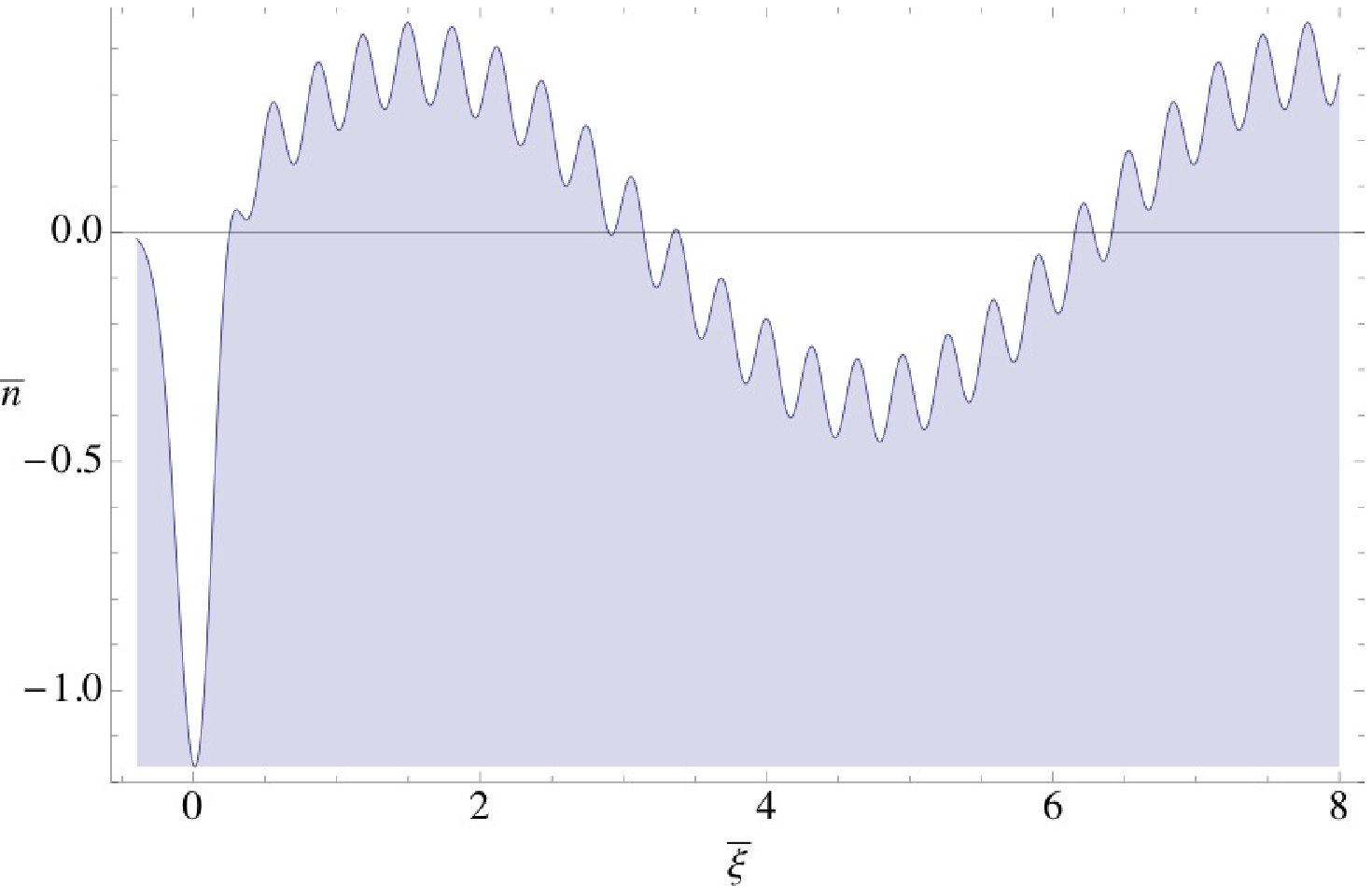}
\caption{The total normalized plasma density wakefield for $n_{0}\simeq
10^{30}\mathrm{cm}^{-3}$ and a Gaussian laser pulse profile $\mathbf{A}%
_{0}^{2}\exp (-\xi ^{2}/L^{2})$with $k_{+}L\simeq 2.1$.}
\end{figure}

\begin{figure}
\includegraphics[width=.9\columnwidth]{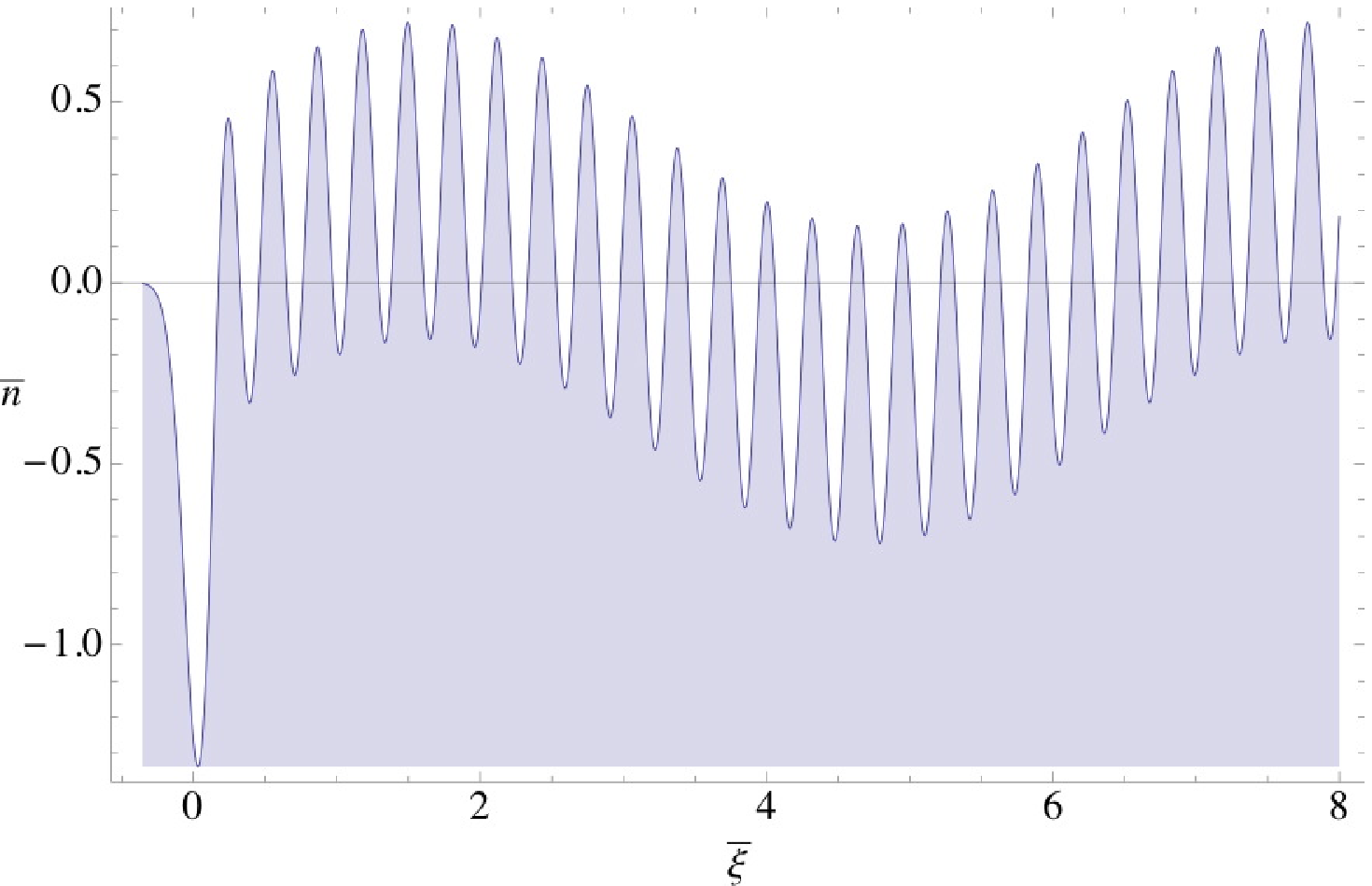}
\caption{The total normalized plasma density wakefield for $n_{0}\simeq
10^{30}\mathrm{cm}^{-3}$ and a Gaussian laser pulse profile $\mathbf{A}%
_{0}^{2}\exp (-\xi ^{2}/L^{2})$with $k_{+}L\simeq 1.6$.}
\end{figure}


\begin{thebibliography}{99}
\bibitem{Tajima-1979} T. Tajima and J. M. Dawson, Phys. Rev. Lett. \textbf{43}, 267 (1979).
\bibitem{Shukla-1986} P. K. Shukla {\it et al.}, Phys. Rep. \textbf{138}, 1 (2006);
M. Marklund and P. K. Shukla, Rev. Mod. Phys. \textbf{78}, 549 (2006).
\bibitem{Mendonca-2001} J. T. Mendon\c{c}a, {\sl Theory of Photon Acceleration}
(Institute of Physics, Bristol, 2001).
\bibitem{Mangels-2004} S. P. D. Mangels {\it et al.}, Nature (London) \textbf{431},
535 (2004); C. G. R. Geddens {\it et al.}, {\it ibid.} \textbf{431}, 538 (2004);
F. Faure {\it et al.}, {\it ibid.} \textbf{431}, 541 (2004).
\bibitem{Leemans-2006} W. P. Leemans {\it et al.}, Nature Phys. \textbf{2}, 696 (2006);
N. H. Maltis {\it et al.}, {\it ibid.} \textbf{2}, 749 (2006).
\bibitem{Joshi-2006} C. Joshi, Sci. Am. \textbf{294}, 22 (2006);
Phys. Plasmas \textbf{14}, 055501 (2007).
\bibitem{Blumenfeld-2007} I. Blumenfeld {\it et al.}, Nature (London) \textbf{445}, 741 (2007);
H. P. Schlenvoigt {\it et al.}, Nature Phys. \textbf{4}, 130 (2008).
\bibitem{Bingham-2007} R. Bingham, Nature (London) \textbf{445}, 721 (2007).
\bibitem{Leemans-2009} W. Leemans and E. Esarey, Phys. Today \textbf{55}, 62 (2009).
\bibitem{Glenzer-2007} S. H. Glenzer {\it et al.}, Phys. Rev. Lett. \textbf{98}, 065007 (2007);
A. L. Kritcher {\it et al.}, Science \textbf{322}, 69 (2008); H. J. Lee {\it et al.},
Phys. Rev. Lett. \textbf{102}, 115001 (2009).
\bibitem{Gardner-1995} C. L. Gardner and C. Ringhofer, Phys. Rev. E \textbf{53}, 157 (1995);
F. Haas {\it et al.}, {\it ibid.} \textbf{62}, 2763 (2000).
\bibitem{Manfredi-2005}  G. Manfredi, Fields Inst. Comm \textbf{46}, 263 (2005).
\bibitem{Shukla-2006}  P. K. Shukla and B. Eliasson, Phys. Rev.  Lett. \textbf{96}, 245001 (2006);
P. K. Shukla, Phys. Lett. A \textbf{352}, 242 (2006).
\bibitem{Marklund-2007}  M. Marklund and G. Brodin, Phys. Rev. Lett. \textbf{98}, 025001 (2007); 
G. Brodin {\it et al.}, {\it ibid.} \textbf{101}, 245002 (2008).
\bibitem{Shukla-2009} P. K. Shukla, Nature Phys. \textbf{5}, 92 (2009).
\bibitem{Shukla-2007}  P. K. Shukla and B. Eliasson, Phys. Rev. Lett.  \textbf{99}, 096401 (2007).
\bibitem{Shaikh-2007} D. Shaikh and P. K. Shukla, Phys. Rev. Lett. \textbf{99}, 125002 (2007).
\bibitem{Harding-2006} A. K. Harding and D. Lai, Rep. Prog. Phys. \textbf{69}, 2631 (2006). 
\end{thebibliography}
\end{document}